\documentclass [11pt]{article}%
\usepackage{amsmath}
\usepackage{amssymb}
\usepackage{epsfig}
\usepackage{latexsym}
\usepackage{amsfonts}
\usepackage{graphicx}%
\usepackage{varioref}
\usepackage{ifthen}
\begin{document}
\parindent 0mm
\setlength{\parskip}{\baselineskip}
\thispagestyle{empty}
\pagenumbering{arabic}
\setcounter{page}{0} 
\mbox{ }
\rightline{UCT-TP-270/07}\\
\rightline{September 2007}\\
\vspace{3.5cm}
\begin{center}
{\Large \bf Electromagnetic   form factors of the $\Delta$(1232) in\\
\vspace{.3cm}
 Dual-Large $N_{c}$ 
 QCD}\footnote{Work supported
in part by the National Research Foundation (South Africa)}\\
\vspace{.5cm}
{\bf C. A. Dominguez, R. R\"{o}ntsch}\\[.5cm]
Institute of Theoretical Physics and Astrophysics\\
University of Cape Town, Rondebosch 7700, South Africa\\[.5cm]
\end{center}
\vspace{.5cm}
\begin{center}
\textbf{Abstract}
\end{center}
The three electromagnetic form factors of the $\Delta(1232)$ resonance, $G^*_M(q^2)$, $G^*_E(q^2)$, and $G^*_C(q^2)$ are obtained in the space-like region using a 
 Dual Resonance Model realization of QCD in the large $N_c$ limit (Dual-$\mbox{QCD}_{\infty}$). Each form factor involves a single free parameter which is fixed by fitting data on $G^*_M(q^2)$, and on the ratios $R_{EM}(q^2)$ and $R_{SM}(q^2)$. Good agreement with experiment is obtained for all three quantities. Results are then used to predict the $q^2$-dependence of the chiral effective-field theory form factors $g_M(q^2)$, $g_E(q^2)$, and $g_C(q^2)$.


\newpage
\bigskip
\noindent

\section{Introduction}

Quantum Chromodynamics (QCD) in the limit of infinite number of colours ($\mbox{QCD}_{\infty}$) \cite{GTH} is known to be solvable, predicting a hadronic spectrum  made of an infinite number  of zero-width resonances \cite{W}. However, the hadronic parameters in this spectrum (masses, couplings, etc.) remain unspecified, so that a model is required to fix them. Some models of this spectrum have been proposed for heavy quark Green's functions \cite{SH}-\cite{CAD1}, as well as for light quark systems \cite{DR}. The infinite number of zero-width resonances of $\mbox{QCD}_{\infty}$ evokes Veneziano's Dual-Resonance model \cite{VEN}, the precursor of string theory. In fact, drawing from this model, a concrete realization of $\mbox{QCD}_{\infty}$ has been proposed, namely
Dual-$\mbox{QCD}_{\infty}$ \cite{CADPI}-\cite{CADN}. In this framework
the masses and couplings in  three-point Green's
functions are  chosen so that form factors are expressed as an Euler  Beta function of the Veneziano type. The asymptotic Regge behaviour of these form factors in the space-like region is power-like, and controlled by a single free parameter of the model. In the time-like region Dual-$\mbox{QCD}_{\infty}$ is not afflicted by unitarity violations known to affect n-point functions ($n \geq \;4$). In fact, three-point functions can be safely unitarized by simply shifting the poles from the real-axis into the second Riemann sheet in the complex energy (squared) plane \cite{CADPI}, \cite{URRU}-\cite{CAD2}. This procedure also provides an estimate of the corrections to  $\mbox{QCD}_{\infty}$ expected from the fact that $N_c=3$. These corrections are small, and of order $\Gamma/M \; \lesssim 10 \%$, where $\Gamma$ and $M$ are a typical width and hadronic resonance mass, respectively.\\

Dual-$\mbox{QCD}_{\infty}$ has been applied with great success to the pion form factor $F_\pi(q^2)$ in both the time-like   and the space-like regions \cite{CADPI},\cite{BRUCH}, as well as to the nucleon form factors in the latter region \cite{CADN}. For the pion one finds excellent agreement with data in the wide range $- q^2 = 0\;-\;10\;\mbox{GeV}^2$, resulting in a chi-squared per degree of freedom $\chi^2_F \simeq 1.1$. A similar  high quality fit is also achieved in the time-like region. In the case of the nucleon, the results for the form factors $F_1(q^2)$ and $F_2(q^2)$, or $G_M(q^2)$ and $G_E(q^2)$, are in very good agreement with data in the available range $- q^2 = 0\;-\;30\;\mbox{GeV}^2$. Furthermore, once the free parameter in each form factor is fixed by the fit, there follows a prediction for the ratio
$G_E(q^2)/G_M(q^2)$. This ratio is in reasonable agreement with polarization transfer data indicating a strong deviation from the so-called scaling law ($G_E(q^2)/G_M(q^2) \simeq const.$). It is important to mention that  if the free parameter of the Dual-$\mbox{QCD}_{\infty}$ form factor is determined from a fit to the high $- q^2$ data, one subsequently achieves agreement all the way down to $q^2 = 0$. The mean-squared radius then becomes a prediction, which e.g. in the case of the pion agrees with data to within one-third of a standard deviation. Clearly, the converse procedure can be applied with equivalent results, i.e. fixing the free parameter from the radius leads to agreement with space-like data everywhere.\\

While Dual-$\mbox{QCD}_{\infty}$ is a particular realization of $\mbox{QCD}_{\infty}$, it can also be viewed as an Extended Vector Meson Dominance model providing corrections to single rho-meson dominance. These corrections arise from the contributions of the radial excitations of the rho-meson. In fact, as is well known, experimentally $F_\pi(q^2)$ falls off with $- q^2$ faster than a monopole, and $G_M(q^2)$ faster than a dipole. In Dual-$\mbox{QCD}_{\infty}$ this is precisely the case, the source of the correction being the radial excitations of the rho-meson. At this point it should be mentioned that Perturbative QCD (PQCD) together with some empirical counting rules \cite{PQCD}
lead to  monopole and dipole  type of asymptotic behaviour for $F_\pi(q^2)$ and $G_M(q^2)$, respectively. However, these results are believed to hold only at extreme asymptotic momenta. In fact, while deep-inelastic scattering indicates precocious scaling, this does not seem to be the case for exclusive processes, such as elastic scattering, or semi-inclusive reactions such as tau-lepton decay into hadrons.\\

In the past few years there has been a renewed interest in understanding the electromagnetic structure of the $\Delta(1232)$, largely motivated by high precision measurements of the photon induced $N \rightarrow \Delta(1232)$ transition at electron beam laboratories (LEGS, BATES, ELSA, MAMI, and J-LAB) \cite{PASCAL}. Concurrently, on the theory sector, lattice QCD and several dynamical models, as well as chiral effective-field theories, have been used to confront the data \cite{PASCAL}. 
Motivated by the success of Dual-$\mbox{QCD}_{\infty}$ in accounting for the data on the pion and nucleon form factors,
we perform here a determination in this framework, and in the space-like region, of $G^*_M(q^2)$, $G^*_E(q^2)$, and $G^*_C(q^2)$, the so-called Jones-Scadron \cite{JS} electromagnetic  form factors of the $\Delta$(1232). After fixing each one of the three free parameters from fits to data, a prediction is obtained for the chiral effective-field theory form factors $g_M(q^2)$, $g_E(q^2)$, and $g_C(q^2)$.\\

\section{Form factors in Dual-$\mbox{QCD}_{\infty}$}

In the literature there are quite a few conventions for the three electromagnetic form factors of $\Delta(1232)$. Some of these are affected by kinematical singularities, and not all are dimensionless. We choose here the Jones-Scadron definition \cite{JS} in which all three form factors are free of kinematical singularities, and are dimensionless.  Generally, in the framework of $\mbox{QCD}_{\infty}$ one expects the form factors to be given by

\begin{equation}
G^*_{i}(s) = \sum_{n=0}^{\infty}
\frac{C_{i n}}{(M_{n}^{2} -s)} \; ,
\end{equation}

where $i=M, E, C$ correspond to $M$ (magnetic dipole), $E$ (electric quadrupole), and $C$ (Coulomb quadrupole), respectively,  $s \equiv q^{2}$, and the masses of the vector-meson zero-width
resonances, $M_n$, as well as their couplings $C_{i n}$ ,
are not predicted in this framework. In Dual-$\mbox{QCD}_{\infty}$ these are chosen so that the form factors are given by  Euler Beta functions, i.e.

\begin{equation}
C_{i n} = G^*_i(0) \;\frac{\Gamma(\beta_{i}-1/2)}{\alpha' \sqrt{\pi}} \;
\frac{(-1)^n} {\Gamma(n+1)} \;
\frac{1}{\Gamma(\beta_{i}-1-n)} \; ,
\end{equation}

where $\beta_{i} (i=M, E, C)$ are free parameters controlling, respectively, the asymptotic behaviour of the form factors $G^*_M(Q^2)$, $G^*_E(Q^2)$, $G^*_C(Q^2)$ in the space-like region ($s<0$), and $\alpha' = 1/2 M^2_{\rho}$ is the universal string tension entering  the rho-meson Regge trajectory

\begin{equation}
\alpha_{\rho}(s) = 1 + \alpha ' (s-M_{\rho}^{2}) \; .
\end{equation}

The mass spectrum is chosen as \cite{AB}

\begin{equation}
M_{n}^{2} = M_{\rho}^{2} (1 + 2 n) \; .
\end{equation}
Using Eqs.(2) and (4) in Eq.(1) one obtains
\begin{eqnarray}
G^*_{i}(s) &=& G^*_i(0) \; \frac{\Gamma(\beta_{i}-1/2)}{\sqrt{\pi}} \;
\sum_{n=0}^{\infty}\;
\frac{(-1)^{n}}{\Gamma(n+1)} \; \frac{1}{\Gamma(\beta_{i}-1-n)} \; 
\frac{1}
{[n+1-\alpha_\rho(s)]} \nonumber \\ [.2cm]
& = & G^*_i(0) \;
\frac{1}{\sqrt{\pi}} \; \frac{\Gamma (\beta_{i}-1/2)}{\Gamma
(\beta_{i}-1)} \;\;
B(\beta_{i} - 1,\; 1/2 - \alpha' s) \;,
\end{eqnarray}

\begin{figure}[ht]
\begin{center}
\includegraphics[height=85mm,width=115mm]{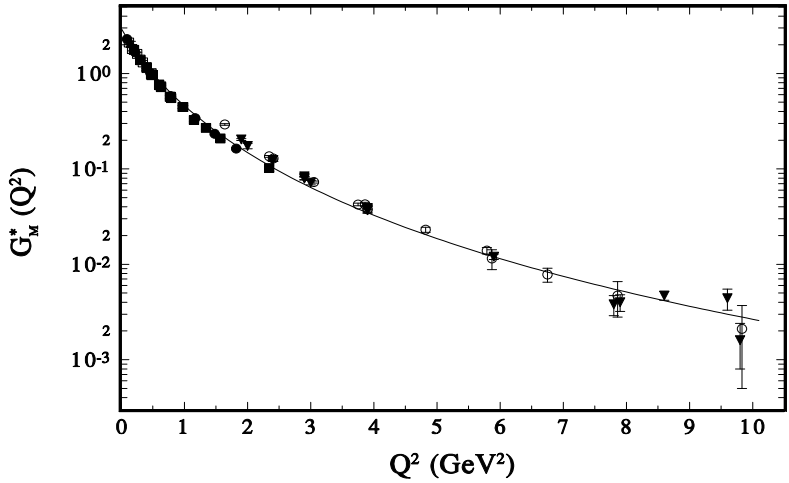}
\caption{The magnetic form factor $G^*_M(Q^2)$, Eq. (5), with $\beta_M = 4.6$, together with the data \cite{DATA1}.}
\end{center}
\end{figure}

where $B(x,y) \equiv \Gamma(x)\; \Gamma(y)/\Gamma(x+y)$ is the Euler Beta function. In the time-like region ($s>0$) the
poles of the Beta function  correspond to an
infinite set of zero-width resonances with equally spaced squared masses given by Eq.(4). In fact, from Eq.(5) it follows

\begin{equation}
Im \; G_{i}(s) = G^*_i(0) \; \frac{\Gamma(\beta_{i}-1/2)}{\alpha' \sqrt{\pi}} \;
\sum_{n=0}^{\infty} \; \frac{(-1)^{n}}{\Gamma(n+1)} \;
 \frac{1}{\Gamma(\beta_{i}-1-n)} \; \pi \; \delta(M_n^2-s) \;.
\end{equation}

Asymptotically, the Regge behaviour of the form factors in the space-like region is power-like, i.e.

\begin{equation}
\lim_{s \rightarrow - \infty} G_{i}(s) = (- \alpha' \;s)^{(1-\beta_{i})}  \; ,
\end{equation}

The free parameters $\beta_{i}$ can be fixed from  fits to the data in this region. Notice that the values $\beta_{i} = 2$ reduce the form factors to  single rho-meson dominance (naive Vector Meson Dominance).\\

The mass formula Eq.(4) predicts e.g. for the first three radial excitations: $M_{\rho'} \simeq 1340$ MeV, $M_{\rho''} \simeq 1720$ MeV,
and $M_{\rho'''} \simeq 2034$ MeV
in reasonable agreement with experiment \cite{PDG} : $M_{\rho'} = 1465 \pm
25$ MeV, $M_{\rho''} = 1700 \pm 20$ MeV,
and $M_{\rho'''} = 2149 \pm 17$ 
MeV. Alternative, e.g. non-linear, mass formulas have been proposed in an attempt to match the asymptotic Regge behaviour to the Operator Product Expansion of current correlators at short distances \cite{ESPRIU}. However, the differences with linear formulas in the values of the first few masses are at the level of a few
percent. Hence, the form factors are hardly modified, since the contribution from higher resonances is factorially suppressed, as seen from Eq.(5).\\

\begin{figure}
\begin{center}
\includegraphics[height=75mm,width=115mm]{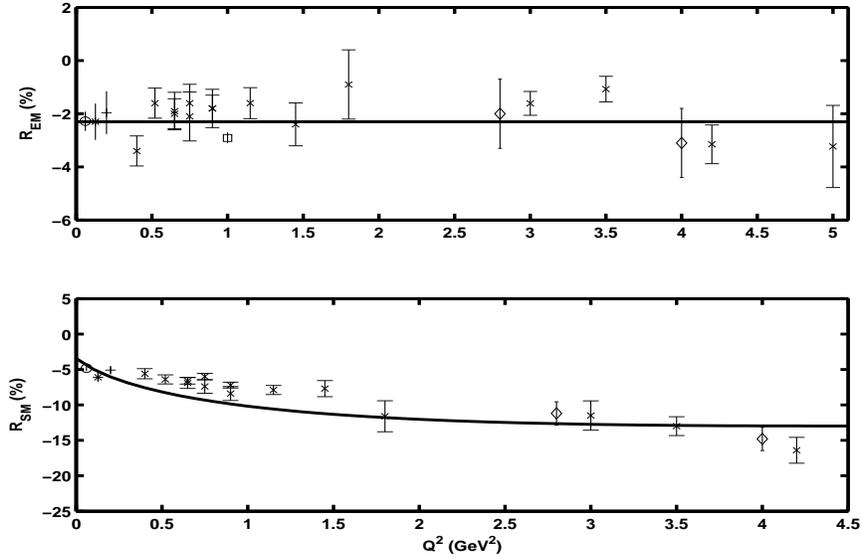}
\caption{The ratios $R_{EM}(Q^2)$ and $R_{SM}(Q^2)$ defined in Eq.(8) with $\beta_M = \beta_E = 4.6$, and $\beta_C = 6.2$, together with the data \cite{DATA2}.}
\end{center}
\end{figure}

\section{Results}
In order to fix the free parameter in each of the three form factors we use data on $G^*_M(q^2)$, and on the ratios

\begin{equation}
R_{EM} = - \;\frac{G^*_E(Q^2)}{G^*_M(Q^2)} \;,\;\;\;\;\; 
R_{SM} = - \;\frac{Q_+ Q_-}{4\;M^2_\Delta} \;\frac{G^*_C(Q^2)}{G^*_M(Q^2)}
\;,
\end{equation}

where
\begin{equation}
Q^2_{\pm} = (M_\Delta \pm M_N)^2 + Q^2 \;,
\end{equation}

and in the sequel we use $Q^2 = - q^2 = - \,s \geq 0$. 
The normalization of the form factors $G^*_i(Q^2)$ is \cite{PASCAL}:
$G^*_M(0) = 3.04$, $G^*_E(0) = 0.07 $, and $G^*_C(0) = 1.00$. The ratios above are then normalized as: $R_{EM}(0) =  - 2.30 \%$, and $R_{SM}(0) = - 3.46 \%$.
The best fit to data on $G^*_M(Q^2)$ is achieved for $\beta_M = 4.6-4.8$. Figure 1 shows  $G^*_M(Q^2)$ for $\beta_M = 4.6$ together with the experimental data \cite{DATA1}. The value $\beta = 4.8$ changes only slightly the high $Q^2$ tail of this curve.
Next, we fit the ratio $R_{EM}(Q^2)$, Eq. (8). Given the large experimental errors in this quantity it is not possible to go beyond the obvious fit $R_{EM}(Q^2) \simeq constant$, or $\beta_M \simeq \beta_E$. This result is shown in Fig.2 for $\beta_M = \beta_E = 4.6$, together with the data \cite {DATA2}. Future improvement in the accuracy of this data  might reveal some $Q^2$-dependence in this ratio, which would require slightly different values of $\beta_E$. 
\begin{figure}
\begin{center}
\includegraphics[height=80mm,width=115mm]{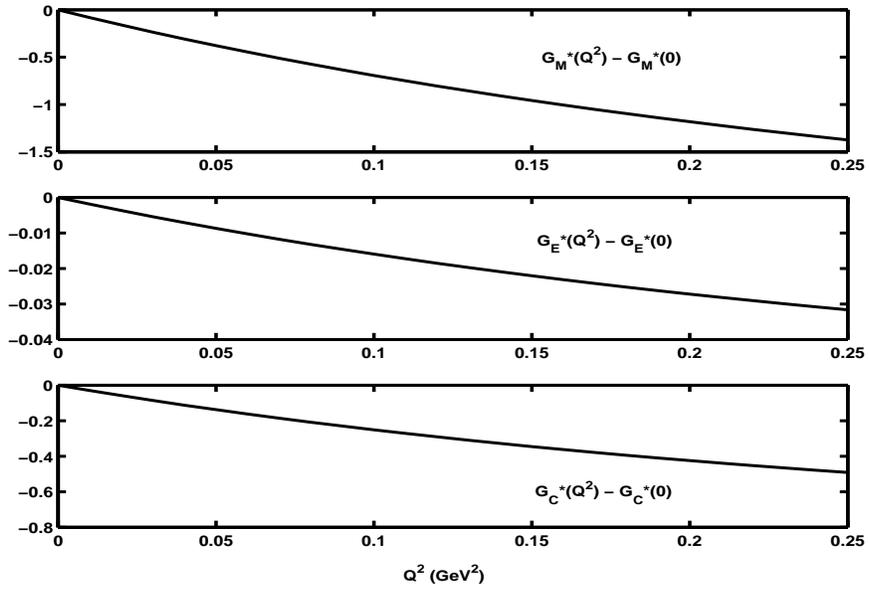}
\caption{The differences $G^*_i(Q^2) - G^*_i(0)$, ($i=M.E,C$), for $\beta_M = \beta_E = 4.6$ and $\beta_C = 6.2$.}
\end{center}
\end{figure}

\begin{figure}[ht]
\begin{center}
\includegraphics[height=80mm,width=115mm]{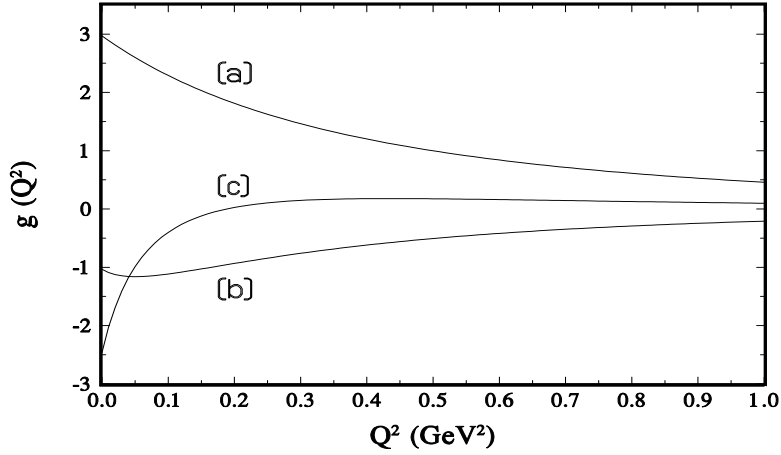}
\caption{The chiral effective-field theory form factors $g_M(Q^2)$ (curve [a]), $g_E(Q^2)$ (curve [b]), and $g_C(Q^2)$ (curve [c]),  at low $Q^2$.}
\end{center}
\end{figure}

\begin{figure}[ht]
\begin{center}
\includegraphics[height=75mm,width=115mm]{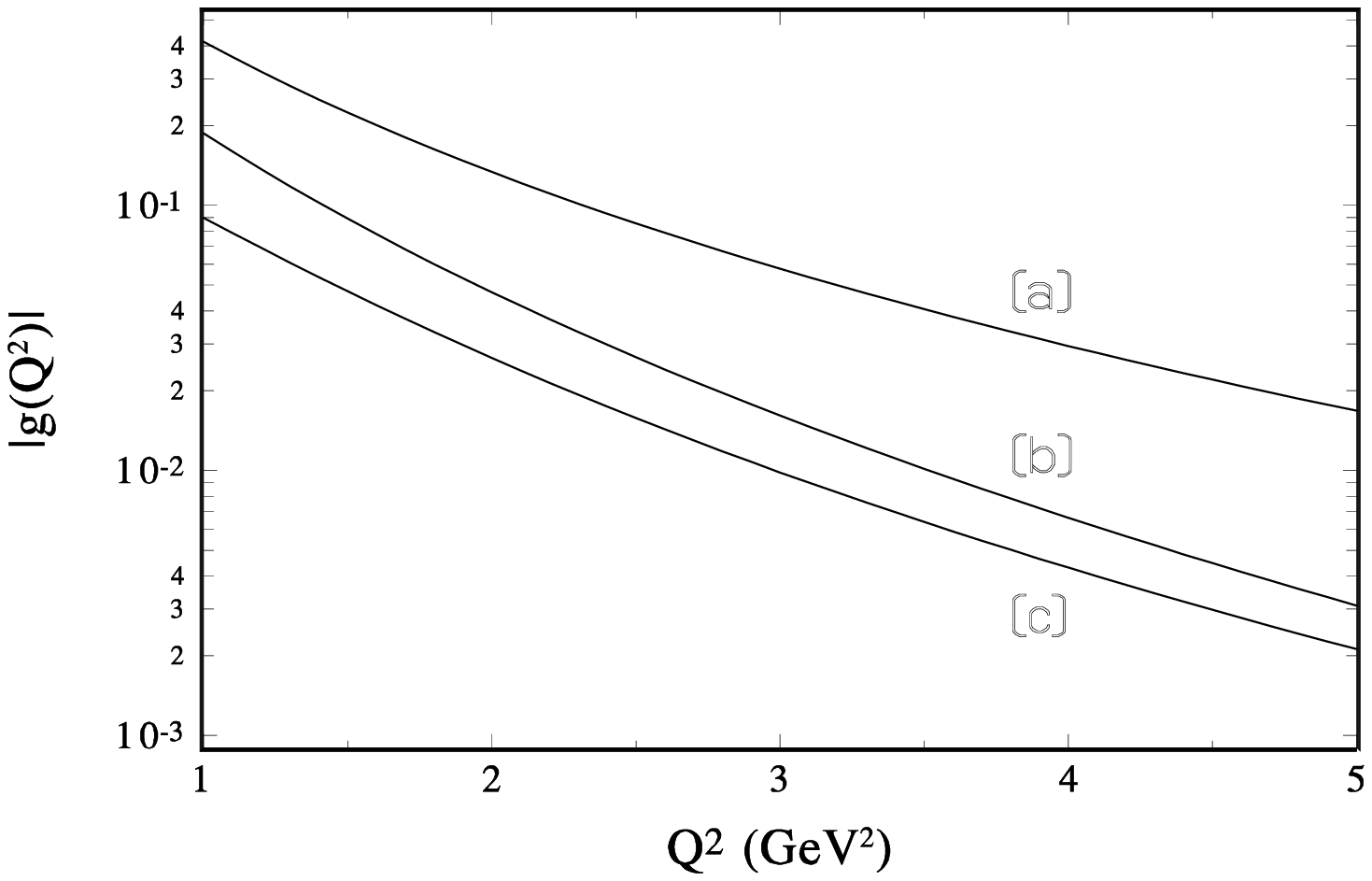}
\caption{The absolute values of the chiral effective-field theory form factors $|g_M(Q^2)|$ (curve [a]), $|g_E(Q^2)|$ (curve [b]), and $|g_C(Q^2)|$ (curve [c]),  at intermediate $Q^2$.}
\end{center}
\end{figure}
Finally, the ratio $R_{SM}(Q^2)$, Eq. (8), can be fitted with $\beta_M = 4.6$ and $\beta_C = 6.0 - 6.2$. This is shown in Fig.2  together with the data \cite{DATA2}.
In Fig. 3 we show the differences $G^*_i(Q^2) - G^*_i(0)$, ($i=M, E, C$), which can be compared with a similar figure given in \cite{PASCAL} showing predictions from several dynamical models \cite{MODELS}-\cite{PASCAL2}. Our results appear to agree qualitatively and quantitatively with most of these models.\\

There is an alternative set of electromagnetic form factors of the $\Delta(1232)$, which are being used in the framework of a chiral effective-field theory \cite{PASCAL2}, namely $g_M(Q^2)$, $g_E(Q^2)$, and $g_C(Q^2)$. These are related to the Jones-Scadron form factors as
\begin{equation}
g_M(Q^2) = G^*_M(Q^2) - G^*_E(Q^2) \;,
\end{equation}

\begin{equation}
g_C(Q^2) = Q^2_+ \; \frac{[ \,\mu(Q^2) \;G^*_C(Q^2) + 4 \;M_\Delta^2\; G^*_E(Q^2)\,]}{\mu^2(Q^2) + 4 \;M_\Delta^2 \;Q^2}
 \;,
\end{equation}

\begin{equation}
g_E(Q^2) = 2 \; \frac{[\, Q_+^2 \;G^*_E(Q^2) - Q^2  \;g_C(Q^2)\,]}{\mu(Q^2)}
 \;,
\end{equation}

where $\mu(Q^2) \equiv M_N^2 - M_\Delta^2 + Q^2$. These form factors are normalized as \cite{PASCAL}: $g_M(0) = 2.97$, $g_E(0) = - 1.00$, and $g_C(0) = - 2.60$. \\
\newpage
Predictions for these form factors are shown in Fig. 4 at low $Q^2$, and as absolute values in Fig. 5 at intermediate $Q^2$, both for $\beta_M = \beta_E = 4.6$ and $\beta_C = 6.2$ in the Jones-Scadron form factors.                   

\section{Conclusions}
Motivated by the good results obtained for the pion and nucleon electromagnetic form factors in  Dual-$\mbox{QCD}_{\infty}$ \cite{CADPI}-\cite{CADN}, this framework has been used here to parametrize the three Jones-Scadron electromagnetic form factors of the $\Delta(1232)$ in the space-like region. These form factors have the advantage of being dimensionless and free of kinematical singularities \cite{JS}.  The single free parameter in each form factor has been fixed by fitting experimental data on $G^*_M(Q^2)$ and on the ratios $R_{EM}(Q^2)$, and $R_{SM}(Q^2)$. Very good agreement with the data is achieved for $G^*_M(Q^2)$ in the wide range of momentum transfers $Q^2 = 0 - 10 \; \mbox{GeV}^2$. For $\beta_M \simeq 4.6$ in Eq.(5), $G^*_M(Q^2)$ falls off with $Q^2$ faster than a dipole. In fact, from Eq.(7) one finds $G^*_M(Q^2) \sim (\alpha' Q^2)^{- 3.6}$. This follows the trend that starts with the pion form factor, which falls off faster than a monopole,  $F_\pi(Q^2) \sim
(\alpha' Q^2)^{- 1.3}$,  and the nucleon form factors, which fall off faster than a dipole,
$F_1(Q^2) \sim (\alpha' Q^2)^{- 2.03}$, $F_2(Q^2) \sim (\alpha' Q^2)^{- 3.2}$. Future data on $G^*_M(Q^2)$ at higher $Q^2$ might require some fine tuning on the value of $\beta_M$. Current data on the ratio $R_{EM}(Q^2)$, needed to fix $\beta_E$, extends only up to $Q^2 \simeq 5 \; \mbox{GeV}^2$. Within the rather large errors, this ratio is consistent with a constant, or $\beta_E \simeq \beta_M$. The ratio $R_{SM}(Q^2)$, though,  does show an appreciable $Q^2$-dependence leading to $\beta_C = 6.2$. The values found for all the three parameters $\beta_i$ carry an uncertainty at the level of a couple of a percent.\\
Having fixed the Jones- Scadron form factors, $G^*_i(Q^2)$, predictions then follow for an alternative set of form factors, $g_i(Q^2)$,  used in a chiral effective-field theory \cite{PASCAL2}. These predictions should be of interest in this framework as they provide simple analytical expressions for the form factors in a wide range of momentum transfers.


\begin{thebibliography}{99}
\bibitem{GTH} G. 't Hooft, Nucl. Phys. B 72 (1974) 461.
 
\bibitem{W} E. Witten, Nucl. Phys. B 79 (1979) 57.
 
\bibitem{SH}  B. Chibisov {\it et al.} Int. J. Mod. Phys. A 12 (1997) 2075;
  B. Blok, M. Shifman and Da-Xin Zhang, Phys. Rev.  D 57 (1998) 2691.
  
\bibitem{CAD1} P. Colangelo, C.A. Dominguez and G. Nardulli,
  Phys. Lett.  B 409 (1997) 417.
 
\bibitem{DR} S. Peris, B. Phily, E. de Rafael, Phys. Rev. Lett. 86 (2001)
  14, and references therein.
 
\bibitem{VEN} P.H. Frampton, {\it Dual Resonance Models}, Benjamin (1974).

\bibitem{CADPI} C.A. Dominguez, Phys. Lett. B 512 (2001) 331.

\bibitem{CADN} C.A. Dominguez, T. Thapedi, J. High Energy Phys. 0410 (2004) 003.

\bibitem{URRU} L.F. Urrutia, Phys. Rev. D 9 (1974) 3213.

\bibitem{CAD2} C.A. Dominguez, Phys. Rev. D 16 (1977) 2320.

\bibitem{BRUCH} C. Bruch, A. Khodjamirian, J.H. K\"{u}hn,
Eur. Phys. J. C 39 (2005) 41.

\bibitem{PQCD} A.V. Efremov, A.V. Radyushkin, Phys. Lett. B 94 (1980) 245; G.P. Lepage, S.J. Brodsky, Phys. Lett. B 83 (1979) 359;  Phys. Rev. D 22 (1980) 2157; A. Duncan, A.H. Mueller, Phys. Rev. D 21 (1980) 1636.

\bibitem{PASCAL} For a comprehensive review see: V. Pascalutsa, M. Vanderhaeghen, S.N. Yang, 
Phys. Rept. 437 (2007) 125.

\bibitem{JS} H.F. Jones, M.D. Scadron, Ann. Phys. (NY) 81 (1973) 1.

 \bibitem{AB} This mass formula was first proposed in a different
  context by A. Bramon, E. Etim, M. Greco, Phys. Lett. 41 B (1972) 609;
  M. Greco, Nucl. Phys. B 63 (1973) 398.
  
 \bibitem{PDG} Review of Particle Physics, Particle Data Group, J. Phys. G: Nucl. Part. Phys. 33 (2006) 1.

\bibitem{ESPRIU} S.S. Afonin, A.A. Andrianov, V.A. Andrianov, D. Espriu,
J. High Energy Phys. 04 (2004) 039. 

\bibitem{DATA1} W. Bartel {\it et al.}, Phys. Lett. B 28 (1968) 148; J. Bleckwen {\it et al.}, DESY Report 71/63 (1971), unpublished; K. B\"{a}tzner {\it et al.}, Phys. Lett. B 39 (1972) 575; J.C. Alder {\it et al.}, Nucl. Phys. B 46 (1972) 573; S.Stein {\it et al.}, Phys. Rev. D 12 (1975) 1884); P.Stoler, Phys. Rept. 226 (1993) 103; L.M. Stuart {\it et al.}, Phys. Rev. D 58 (1998) 032003.

\bibitem{DATA2} V.V. Frolov {\it et al.}, Phys. Rev. Lett. 82 (1999) 45; 
R.Beck {\it et al.}, Phys. Rev. C 61 (2000) 035204; K. Joo {\it et al.}, Phys. Rev. Lett. 88 (2002) 122001;
 N.F. Sparveris, Phys. Rev. Lett. 94 (2005) 022003; J.J. Kelly {\it et al.}, Phys. Rev. Lett. 95 (2005) 102001;
M. Ungaro {\it et al.},  Phys. Rev. Lett. 97 (2006) 112003; S. Stave {\it et al.}, Eur. Phys. J. A 30 (2006) 471.

\bibitem{MODELS} T. Sato, T.-S.H. Lee, Phys. Rev. C 54 (1996) 2660; {\it ibid} 63 (2001) 055201.

\bibitem{PASCAL2} V. Pascalutsa, M. Vanderhaeghen, Phys. Rev. Lett. 95 (2005) 232001; Phys. Rev. D 73 (2006) 034003. 

\end{thebibliography}
\end{document}